\documentclass[journal=jctcce,layout=twocolumn,manuscript=article]{achemso}

\makeatletter
\let\l@addto@macro\relax
\makeatother
\usepackage[fontsize=10pt]{scrextend}
\usepackage{amsmath}
\usepackage[utf8]{inputenc}
\usepackage{graphicx}
\title{Pressure Profile Calculation with Mesh Ewald Methods}
\author{Marcello Sega}
\affiliation{Computational Physics Group, University of Vienna, Sensengasse 8/9, 1170 Vienna, Austria}
\email{marcello.sega@univie.ac.at}
\author{Balázs Fábián}
\affiliation{Institut UTINAM (CNRS UMR 6213), Université de Franche-Comté, 16 route de Gray, F-25030 Besançon, France}
\alsoaffiliation{Department of Inorganic and Analytical Chemistry, Budapest University of Technology and Economics, Szt. Gellért tér 4, H-1111 Budapest, Hungary}
\author{P{\'a}l Jedlovszky}
\affiliation{EKF Department of Chemistry, Leányka u. 6, H-3300 Eger, Hungary}
\alsoaffiliation{Laboratory of Interfaces and Nanosize Systems, Institute of Chemistry, Eötvös Loránd University, Pázmány P. Stny 1/A, H-1117 Budapest, Hungary}
\alsoaffiliation{MTA-BME Research Group of Technical Analytical Chemistry, Szt. Gellért tér 4, H-1111 Budapest, Hungary}

\let\oldmaketitle\maketitle
\let\maketitle\relax

\begin{document}
\twocolumn[
\begin{@twocolumnfalse}
\oldmaketitle
\begin{abstract}
The importance of calculating pressure profiles across liquid
interfaces is increasingly gaining recognition, and efficient methods
for the calculation of long-range contributions are fundamental in
addressing systems with a large number of charges.  Here, we show how
to compute the local pressure contribution for
mesh-based Ewald methods, retaining the typical $N \log N$ scaling
as a function of the lattice nodes $N$. This is a considerable
improvement on existing methods, which include approximating the
electrostatic contribution using a large cut-off and the, much
slower, Ewald calculation.  As an application, we calculate the
contribution to the pressure profile across the water/vapour
interface, coming from different molecular layers, both including
and removing the effect of thermal capillary waves.  We compare the
total pressure profile with the one obtained using the cutoff
approximation for the calculation of the stresses, showing that the
stress distribution obtained by the Harasima and Irving-Kirkwood
are quite similar and shifted with respect to each other at most
0.05~nm.\\
\end{abstract}
\end{@twocolumnfalse}
]

\section{Introduction} 
\begin{figure}[t]
\includegraphics[width=\columnwidth,trim=50 200 50 -100, clip ]{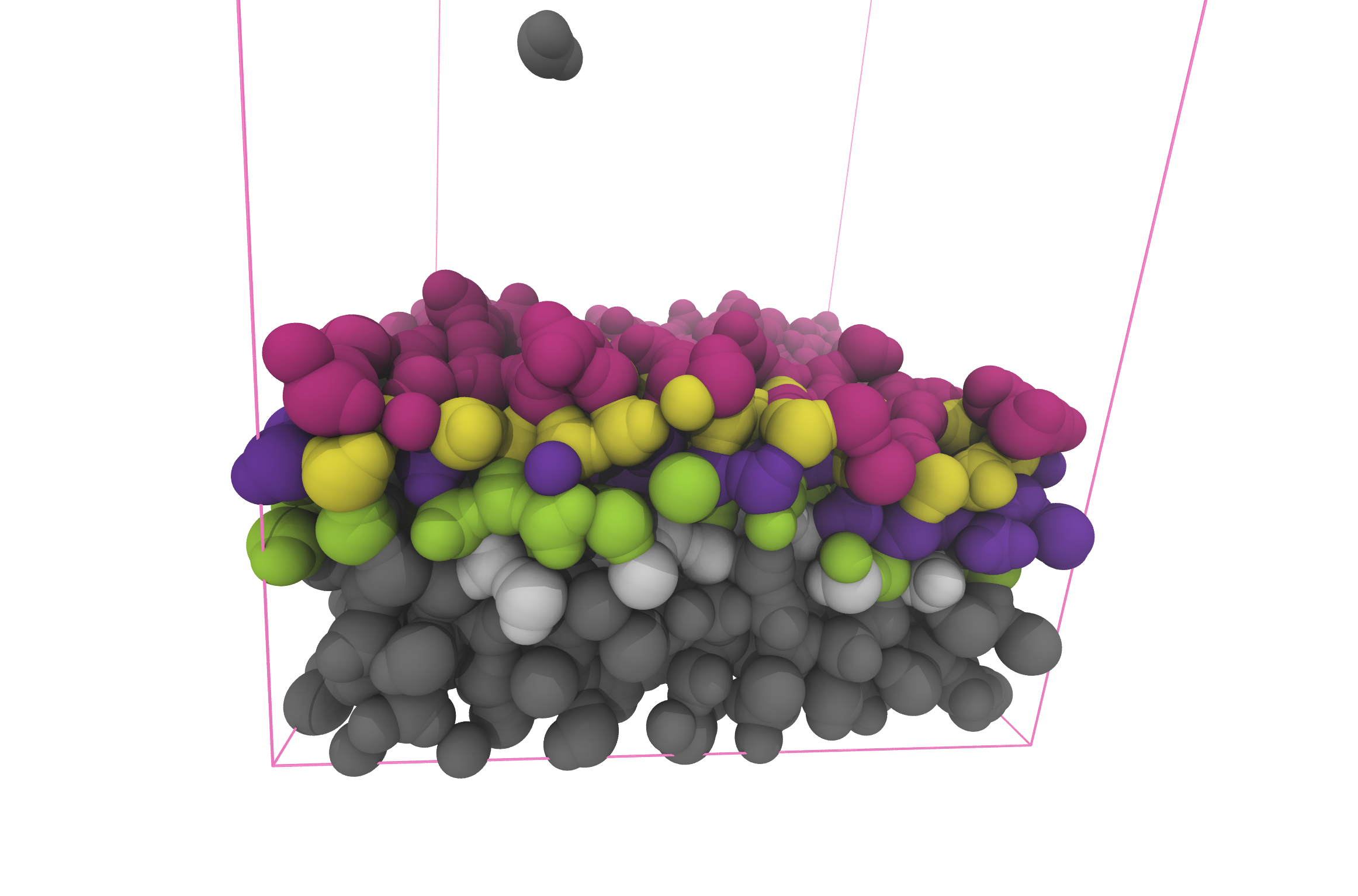}
\caption{Simulation snapshot of a water/vapour interface. The first five molecular layers in the upper half of the simulation box (lower half not shown) are highlighted with different colors. One molecule identified to be in the vapour phase by a cluster-search algorithm can be also seen above the first layer of the surface\label{snap}}
\end{figure}

Calculating the profiles of various physical
quantities such as mass, charge, or electrostatic potential, is a
standard approach used in computer simulations in order to characterize
inhomogeneous systems including fluid (soft) interfaces, like the water/vapour one depicted in Fig.\ref{snap}, membranes,
micelles or other self-associates. While some of these profiles have
been routinely calculated in the past decades, the importance of
determining pressure profiles is only recently being recognized,
for example, for the calculation of the mechanical properties of
macromolecules \cite{schulten,ollila20093d,hatch2012molecular,torres}.
Conjectures related to pressure profiles also play a key role in the
possible explanations of several phenomena. Thus, for instance, it
was proposed by Cantor that the molecular mechanism of anesthesia
is related to the alteration of the lateral pressure profile inside
the cell membrane\cite{cantor1997lateral}; Imre et al.~claimed that
the spinodal pressure of a liquid phase can be extracted from the
lateral pressure profile obtained at the liquid-vapour interface of
the same system at the same temperature\cite{imre2008estimation}.
Experimental verification or falsification of these conjectures is
an extremely challenging task as it requires the measurement of
the mean lateral pressure across interfaces, or macromolecules at
atomic resolution, a task that is rarely accomplished (see, for
example Ref.~\cite{templer}).

The calculation of pressure profiles in computer simulations, on
the other hand, is complicated by the fact that it requires the
local determination of a quantity that is inherently non-local.
More precisely, a local pressure tensor cannot, in general, be
uniquely defined. Instead, it can only be given up to a divergence-free
second rank tensor in a path dependent form. Several integration
paths, such as the Irving-Kirkwood\cite{irving1950statistical} and Harasima\cite{harasima1958molecular} paths,
were shown to provide comparable profiles\cite{schofield1982statistical}. Besides its conceptual
simplicity, the Irving-Kirkwood path has the advantage of leading
to a formula for the local pressure that gives access to both the
normal and lateral components of the virial, distributed homogeneously
along the paths connecting pairs of particles.  On the other hand,
while the Harasima path allows one to calculate only the lateral
components of the pressure tensor, these are concentrated at the
particles' positions. In other words,
a certain amount of the total lateral pressure is associated with each particle, making the calculation
of the lateral pressure profile straightforward and computationally
efficient\cite{sonne,sega2015layer}.

Another serious technical problem in calculating
pressure profiles in charged systems is how to take into account the
long range correction term of the electrostatic interaction. It is
now well-established that neglecting this correction to the energy
and forces can lead to important systematic errors. The method of
reaction field correction\cite{onsager1936electric}, which uses the dielectric constant
of the continuum beyond a suitably chosen cutoff sphere, is
unsuitable for strongly inhomogeneous systems. The Ewald summation\cite{ewald} can, in principle, be
used in such systems; the contribution of its correction term
to the pressure profile was derived by Sonne and coworkers for the
Harasima path\cite{sonne}.  They also showed that potentials that
are not strictly pairwise additive, such as the correction term of
the Ewald summation, cannot be used in combination with the
Irving-Kirkwood profile. However, because of its poor scaling
properties, the use of the Ewald summation is prohibitive for systems
with a large number of charges.

In theory, this problem can be overcome
using a method based on the Fast Fourier Transform (FFT), such as
mesh Ewald methods\cite{darden1993particle,essmann1995smooth,hockney1988computer}.
However, the local pressure from the reciprocal space contribution of  mesh Ewald methods has not, to the best of our knowledge, been derived yet.

For approaches based on the Irving-Kirkwood path, as it is not
possible to take into account non-pair additive potentials,
several authors have opted for running the simulations using their
method of choice for the evaluation of long range contributions,
but considering cutoff electrostatics for the purpose of evaluating
the local pressure (see, for example,
Refs.\cite{schulten,kasson,ollila2007polyunsaturation}). This is
usually done by re-running the simulation and using cutoff values
that are larger than usual, thus reducing the artifacts. However, in order to reach a
precision of the order of 1 bar, cutoff values larger than 2.5~nm
must be used (see Fig.~7 in Ref.~\cite{vanegas}). Sonne and
collaborators~\cite{sonne} showed that the pressure profile calculated
using the Irving--Kirkwood approach and a cutoff radius of 2 nm is
qualitatively similar to the one calculated using the Harasima path
with long-range electrostatic contributions, although with an
indetermination of roughly 100 to 200~bar, which makes it only a
qualitative test.

In this paper we show how the correction term of the smooth Particle
Mesh Ewald\cite{essmann1995smooth} (sPME) method can be taken into
account in calculating lateral pressure profiles using the Harasima
path in an efficient way that does not alter the scaling of the
long-range correction calculation.  We describe the method itself
in Sec.~\ref{theory} and compare the pressure contributions
associated to particles with those obtained with a reference, simple
Ewald calculation. In Sec.~\ref{example} we apply the method to
the case of the water/vapour interface, and in Sec.~\ref{sec:scaling}
we describe the scaling properties of the algorithm.  Finally, in
Sec.~\ref{conclusions}, we summarize the results presented in this
article.


\section{The reciprocal space contribution\label{theory}}
We start by decomposing the instantaneous pressure tensor $\mathbf{P}$, as customary, in the ideal gas and virial ($\mathbf{\Xi}$) contributions,
\begin{equation}
\mathbf{P} V = \sum_i m_i \mathbf{v}_i \otimes \mathbf{v}_i - 2 \mathbf{\Xi}
\end{equation}
The difference between lateral and normal components of the pressure tensor plays an important role for interfaces, as it describes the surface tension, and its local variant can be written as
\begin{equation}
\gamma(z) = P_\mathrm{N} - P_\mathrm{T}(z),\label{eq:st}
\end{equation}
where we have chosen $P_\mathrm{N}=P_{zz}$ and $P_\mathrm{T} = (P_{xx}+P_{yy})/2$.
This formula applies only to planar interfaces and we will assume, as it is customary, that it applies also in the presence of capillary waves, as long as the surface remains macroscopically flat. The presence of curved interfaces introduces additional difficulties in the definition and calculation of the surface tension\cite{rowlinsonwidom,schofield1982statistical,sodt2012tension}.
Here we note that while $P_\mathrm{N}$ is constant through the system to
ensure mechanical stability\cite{rowlinsonwidom}, $\Xi_{zz}$ is not.

In the Ewald sum, the reciprocal space contributions to energy and virial are expressed as sums over the reciprocal lattice vectors $\mathbf{m}$
\begin{align}
U^\mathrm{rec}     & =    \sum_{\mathbf{m}\neq0} \frac{1}{2\pi V} f(\mathbf{m}^2)  S(\mathbf{-m}) S(\mathbf{m})\nonumber\\
\Xi^\mathrm{rec}_{\mu\nu} & = \frac{1}{V} \sum_{\mathbf{m}\neq0} \frac{1}{2\pi V}  f(\mathbf{m}^2) g_{\mu\nu}(\mathbf{m}) S(-\mathbf{m})S(\mathbf{m})
\label{eq:reciprocal}
\end{align}
where the functions $f$ and $g_{\mu\nu}$ are defined as
\begin{align}
f(\mathbf{m}^2) &= \frac{e^{-\pi^2\mathbf{m}^2/\beta^2}}{\mathbf{m}^2} \\
g_{\mu\nu}(\mathbf{m}) &= \left(\delta_{\mu\nu}-2 \frac{1+\pi^2\mathbf{m}^2/\beta}{\mathbf{m}^2} \mathbf{m}_\mu\mathbf{m}_\nu\right),
\end{align}
and the structure factor is
\begin{equation}
S(\mathbf{m}) = \sum_i q_i  e^{2\pi i \mathbf{m}\cdot\mathbf{r}_i}.
\end{equation}
In mesh-based algorithms, the reciprocal space term in the Ewald
sum is calculated with the aid of the FFT\cite{cooley1965algorithm}
by replacing the point charges, located in the continuum at positions $\mathbf{r}_i$, with a
discretized distribution on a lattice with nodes at positions $\mathbf{r}_p$
\begin{equation}
\tilde{\rho}(\mathbf{r}_p) = \frac{1}{h^3} \sum_i W(\mathbf{r}_p-\mathbf{r}_i),
\end{equation}
where $h$ is the lattice spacing and $W(\mathbf{r}_p-\mathbf{r}_i)$
is a suitable charge assignment function. The structure factor
$S(\mathbf{m})$ is then replaced in Eqs.(\ref{eq:reciprocal}) by its
estimate $\tilde{S}(\mathbf{m}) = B(\mathbf{m})\mathrm{FFT}[\tilde{\rho}]$,
which is now evaluated on the points of the reciprocal lattice, and
where the function $B(\mathbf{m})$ depends on the specific interpolation
scheme\cite{essmann1995smooth,deserno1998mesh}.

The choice of the assignment function for the discrete charge
distribution is not unique, a fact that has led to the appearance of families of
methods such as the
Particle-Particle-Particle-Mesh (P3M) of Hockney and Eastwood\cite{hockney1988computer}, and the
Particle Mesh Ewald (PME) in its original
variant\cite{darden1993particle} that uses Lagrange interpolation or in its ``smooth''
version\cite{essmann1995smooth} (sPME) based on cardinal B-splines. Once a physical quantity pertaining to single particles has been calculated on the lattice, it has to be interpolated back from the lattice to the real position of the particles. This is usually done by the same assignment function used to generate the distribution on the lattice\cite{deserno1998mesh}. If a function $A(\mathbf{r}_p)$ is known at the lattice nodes, it can be distributed back to the atomic positions using
\begin{equation}
 A(\mathbf{r}_i) = \sum_{\mathbf{r}_p} A(\mathbf{r}_p) W(\mathbf{r}_i - \mathbf{r}_p). \label{eq:backinterpol}
\end{equation}

The reciprocal space term of the virial in the Ewald sum can be easily written
in a form that is suitable for the Harasima path formulation, which
associates to particle $i$ the contribution\cite{sonne}
\begin{equation}
\Xi^\mathrm{rec,i}_{\mu\nu}  = \frac{q_i}{V}\sum_{\mathbf{m}\neq0} \mathrm{Re}\left\{\frac{e^{-i\mathbf{m}\cdot\mathbf{r}_i}}{2\pi V}  S(\mathbf{m}) \right\} f(\mathbf{m}^2) g_{\mu\nu}(\mathbf{m}).\label{eq:virial-harasima}
\end{equation}
As we are interested in the diagonal elements of the tensor only, $\mu=\nu$, the real part operator in Eq.(\ref{eq:virial-harasima}) is superfluous because the $f$ and  $g_{\mu\nu}$ are even functions of $\mathbf{m}$.
Another approach has been previously used in literature to take into
account the reciprocal space contribution to the local
pressure for the plain Ewald method\cite{alejandre95a}, but consisted of distributing the
reciprocal space virial contribution equally to each particle, which
is not a justified assumption, as the reciprocal
space force contribution is different for each particle.

Note that, because the contribution in reciprocal space is not
pairwise additive, it is not possible to use an Irving-Kirkwood
formulation either for the Ewald sum, or for its mesh-based
approximations\cite{sonne}.

We now proceed to extending the reciprocal space contribution to mesh-based algorithms, by noting that Eq.(\ref{eq:virial-harasima}) is, for $\mu=\nu$, the discrete inverse Fourier transform of the complex function $S(\mathbf{m})f(\mathbf{m}^2)g_{\mu\nu}(\mathbf{m})$, evaluated at the particles' positions, and scaled by the factor $q_i$. On the lattice, this expression is replaced by the inverse FFT of the function $\tilde{S}(\mathbf{m})f(\mathbf{m}^2)g_{\mu\nu}(\mathbf{m})$. This quantity is eventually interpolated back to the real positions using Eq.(\ref{eq:backinterpol}), leading to the approximate expression for the reciprocal space contribution

\begin{equation}
\tilde{\Xi}^\mathrm{rec,i}_{\mu\mu} = \frac{q_i}{V} \sum_{\mathrm{r}_p}\tilde{\Xi}^\mathrm{rec}_{\mu\mu}(\mathrm{r}_p)W(\mathbf{r}_i - \mathbf{r}_p),\\
\end{equation}
with
\begin{equation}
\tilde{\Xi}^\mathrm{rec}_{\mu\mu} (\mathrm{r}_p)= \mathrm{FFT}^{-1} \left[ B(\mathbf{m}) \mathrm{FFT}[\tilde{\rho}] f(\mathbf{m}^2)g_{\mu\mu}(\mathbf{m})\right].
\label{eq:fft}
\end{equation}

Notice that in order to be consistent with the sum
Eq.(\ref{eq:reciprocal}), the argument of the inverse FFT in
Eq.(\ref{eq:fft}) for $m=0$ must be set explicitly to zero.  The correction
and exclusion terms in real space\cite{essmann1995smooth} are expressed as sum
of pair contributions and, thus, must be taken into account
similarly to other pairwise, short range interactions.

We would like to stress once more that, thanks to the use of the
Harasima formulation, it is possible to associate a virial contribution
to each particle, and the local stresses are thus distributed in
the continuum. In the present case, a grid is used only for the
mesh-Ewald charge spreading procedure and, once the back-interpolation
is performed, the reciprocal space contribution is associated
to each of the particles.

We implemented this approach in our development version (available
free of charge at
\texttt{github.com/Marcello-Sega/gromacs/tree/virial}) of the
popular GROMACS molecular simulation
package\cite{berendsen1995gromacs,pall,sega2015layer}, version 5.0,
which extends the trajectory files (usually containing positions,
velocities and forces) to include also the local virial components
$\Xi_{xx}, \Xi_{yy}$ and $\Xi_{zz}$.

\begin{figure}[t]
\includegraphics[width=\columnwidth,trim=0 0 20 10, clip]{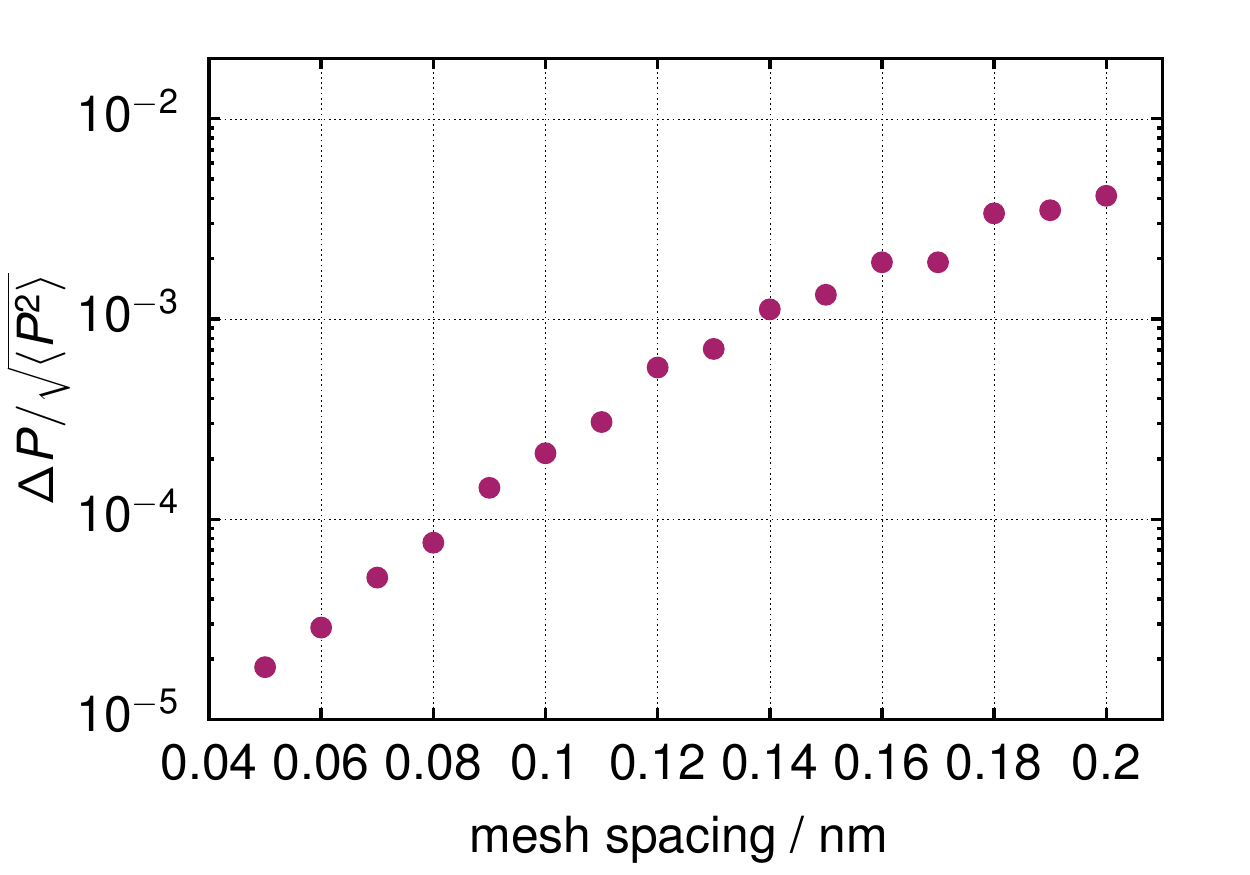}
\caption{Root mean square error per particle for the pressure, as a function of the mesh spacing (charge interpolation order set to 4).\label{accu1}}
\end{figure}

\begin{figure}[t]
\includegraphics[width=\columnwidth,trim=0 0 20 10, clip]{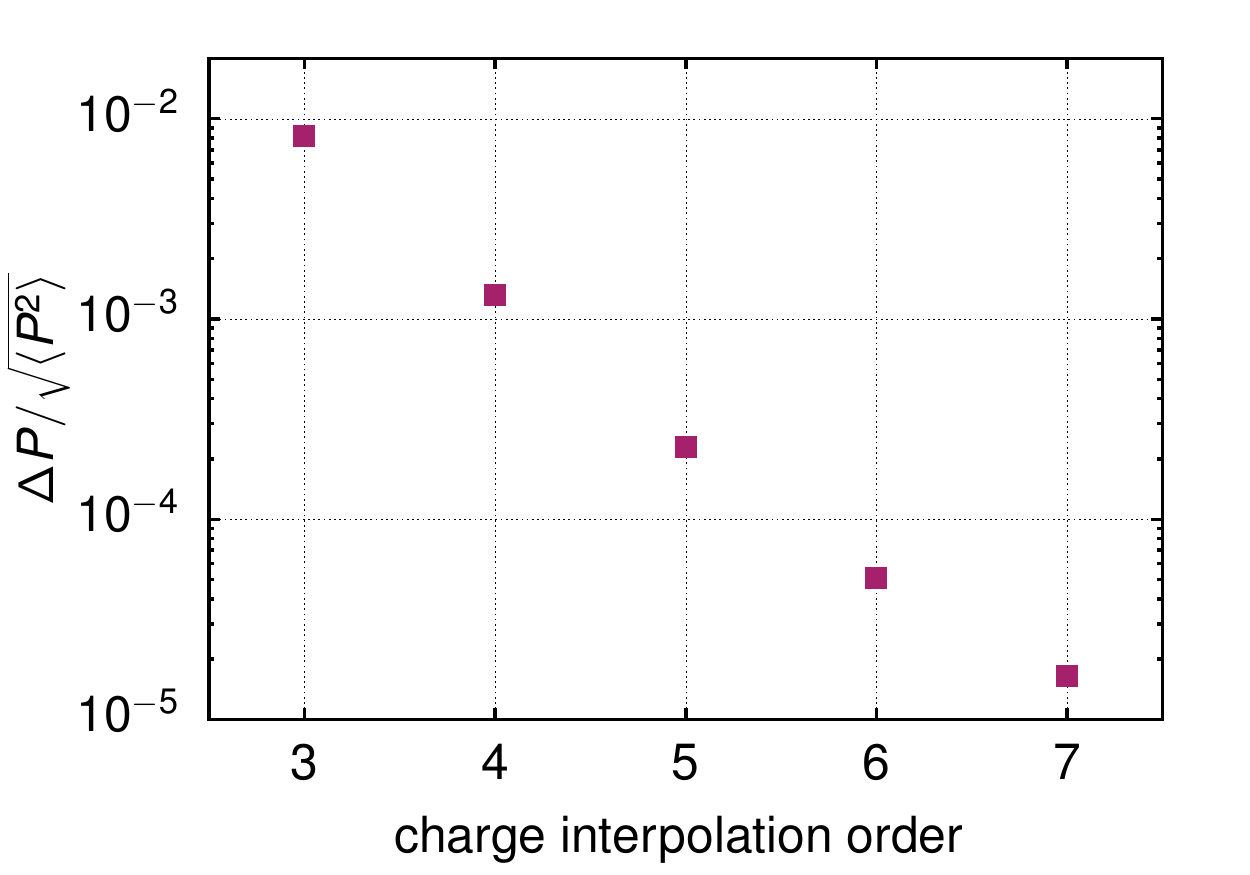}
\caption{Root mean square error per particle for the pressure, as a function of the charge interpolation order (mesh spacing set to 0.15 nm).\label{accu2}}
\end{figure}

In order to check for the correctness and proper implementation of
the algorithm, we tested it on a system of point charges (in absence
of any other type of interaction but the electrostatic one).  As a
first step, we checked that the sum of the pressure terms associated
to the atoms indeed yields the global virial up to roundoff errors.
Subsequently, we computed the root mean square error $\Delta P$
per particle, following Holm and Deserno\cite{deserno1998mesh}, as
\begin{equation}\Delta P = \sqrt{\frac{1}{N}\sum_i \sum_ \mu \left(P_{\mu\mu}^i
-\hat{P}^i_{\mu\mu} \right)^2},
\end{equation} where $N$ is the number of charges
and $\hat{P}_{\mu\mu}^i$ is a reference value, namely the pressure
contribution of particle $i$ calculated on the same configurations using the plain Ewald method,
with a large number of reciprocal space contributions (12 along $x$
and $y$ direction, 36 along $z$ direction), and a relative accuracy
at the cutoff (1.5~nm) of $10^{-5}$.
In Fig.~\ref{accu1} and Fig.~\ref{accu2} we report, respectively, the relative root mean square
error $\Delta P/\sqrt{\langle P^2\rangle}$ per particle, as a function of the mesh spacing and of the
charge interpolation order, where $\langle P^2\rangle = \frac{1}{N}\sum_i\sum_\mu { \hat{P}^{i2}_{\mu\mu}}$.

The average error that is made in estimating the pressure
contribution is less than $0.5\%$ for all choices of parameters and
decreases monotonically with decreasing mesh spacing and increasing
interpolation order, down to $0.001\%$. Similarly, we computed
the root means square relative error on the pressure per particle introduced
by using a simple cut-off scheme, of course always calculated on
the same set of configurations.  If the electrostatic force is
truncated at 1.5 nm, the root mean square error on the
pressure per particle is about $200\%$.

\section{The pressure profile of water/vapour interface\label{example}}
We proceeded to run a simulation of the water/vapour interface,
modeling water molecules with the SPC/E\cite{berendsen1987missing}
potential. While other models compare better with several experimentally known
quantities\cite{vega2006vapor,alejandre2010surface}, the SPC/E
model represents an optimal choice for testing new methods, thanks
to its modest computational requirements.

The initial configuration was generated simply by
increasing the $z$ box edge of a $5\times{}5\times{}5$ nm$^3$
equilibrated water simulation box to 15 nm. To integrate the equations
of motion we used an integration step of 1 fs, constraining all
bonds by means of the SHAKE algorithm\cite{ryckaert1977numerical}
(the local pressure contribution of the constraints were being also taken
into account), and keeping the temperature at the constant value
of 300~K by means of the Nos\'e--Hoover thermostat\cite{nose84a,hoover85a}.
The short-range part of the potentials were cut off either at 1.5 or at 2.0 nm and, at that distance, the relative accuracy of
the direct space contribution to the electrostatic potential was
set to $10^{-5}$. The long-range correction to the electrostatic
potential was calculated using sPME for all trajectories, with a
target lattice spacing of 0.15 nm. No long-range corrections were
used for the van der Waals potential. Configurations, including
atomic positions and virial contributions, were saved to disk every
0.1~ps. The contributions to the lateral virial stemming from the Lennard-Jones interaction
and from constraint forces\cite{kasson} were taken into account as usual with the expression of the virial for the Harasima path
$\Xi^i_{\mu\mu}= - 1/2 \sum_{j\neq{}i}f^{ij}_\mu r^{ij}_\mu $ , where $f^{ij}_\mu$ and $r^{ij}_\mu$ are the components of the pair forces and of the vectors connecting the two interacting atoms, respectively.

\begin{figure}[t]
\includegraphics[width=\columnwidth,trim=0 2 20 0,clip]{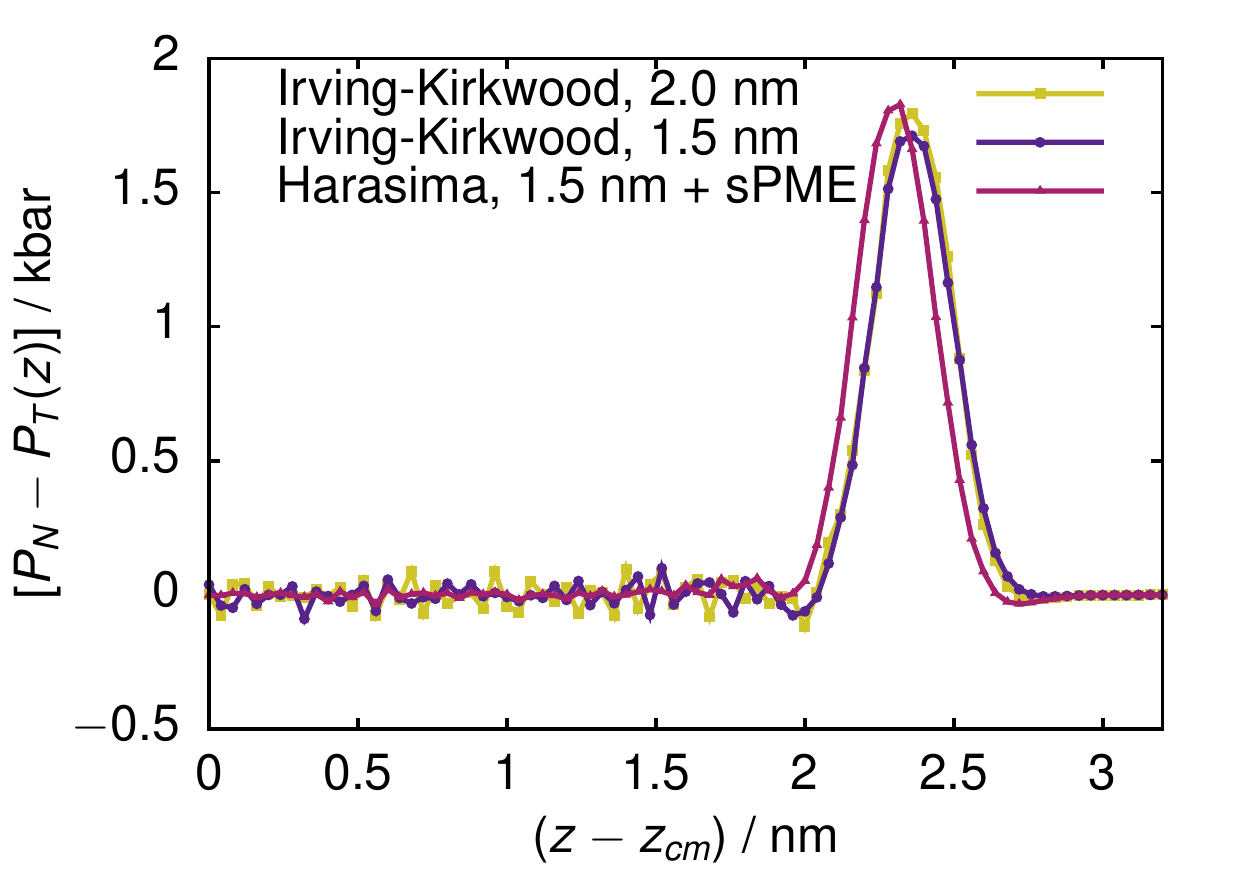}
\label{comparison}
\caption{Comparison between the surface tension profile $\gamma(z)$ obtained with two different paths (Irving-Kirkwood and Harasima) and two different cut-off values for the short-range forces and (in case of the Irving-Kirkwood path) for the calculation of the electrostatic contribution to the pressure. The profile is the average of the two halves of the simulation box, and the origin is in the middle of the water slab. \label{comparison}}
\end{figure}

We calculated the surface tension profile with the Harasima path,
including the full sPME contributions and Lennard-Jones interactions
cut off at $r_c=1.5$ nm, and compared it to the profile calculated
using the procedure of Vanegas  and coworkers\cite{vanegas} using
the Irving-Kirkwood path and cut-off values of 1.5 and 2.0 nm,
respectively. The resulting profiles are reported in Fig.\ref{comparison}.
The two profiles calculated using the Irving-Kirkwood path are
qualitatively similar, differing only in the peak height. On the
other hand, the profile calculated with the Harasima path is shifted
by about 0.05 nm towards the center of the fluid slab. From the
surface tension profiles $\gamma(z)$ it is possible to calculate
both the surface tension $\gamma=(1/2)\int{\gamma(z)}dz$ as well
as the position of the surface of tension\cite{rowlinsonwidom}
$z_s=(1/2\gamma) \int{\gamma(z) |z| dz }$. The values of the surface
tension $\gamma$ as computed from the integral of the
profile (thus neglecting the long-range contributions in the
Irving-Kirkwood case) and from the global pressure tensor are
reported in Tab.~1, together with the values of $z_s$ and of its
distance $\delta=z_e-z_s$ (the Tolman length) from the location of the equimolar surface
$z_e$. The equimolar surface is defined as the one that divides a volume $V$ in two regions (liquid and vapour) of volumes $V_l$ and $V_v$, respectively, such that $\rho V = \rho_v V_v + \rho_l V_l$, with $\rho_v$ and $\rho_l$ the densities of the two phases as measured far from the interface\cite{rowlinsonwidom}. Since we have two interfaces, we apply this criterion to half of the simulation box (the liquid phase being centerd in the box), and obtain that $\rho_l z_e + \rho_v (L/2 -z_e ) = \rho L/2$, or $z_e = L (\rho- \rho_v)/ 2(\rho_l -\rho_v)$. Since in our case $\rho_v/\rho \simeq \rho_v/\rho_l \simeq 10^{-5}$, we can safely use the approximation $z_e\simeq\rho L / (2\rho_l)$.

The equimolar surface was found to be located at  $z_e=2.43\pm0.01$ and $2.40\pm 0.01$ nm for the 1.5 and 2.0 nm cutoff cases, respectively.

For the sake of comparison, in Tab.~1 we report some data from the literature, obtained with different mesh parameters and cutoff treatment, for the surface tension of the SPC/E model at 300 K. The values reported here are those which do not include analytical tail corrections, as in the present work.

It is interesting to note that analytical results estimate the value of $\delta$ to be in the range of the molecular size $r_m$ (at least, for pair potentials). More precisely, $\delta\simeq r_m/3$ and $r_m/4$ for the Irving-Kirkwood and Harasima paths, respectively\cite{rowlinsonwidom}. This means, in the case of water, $\delta \simeq 0.1$ and 0.08 nm for the two paths, respectively. These values are not incompatible with those obtained, to the best of our knowledge for the first time in this work, by numerical simulation.

\begin{table}[t]
\begin{tabular}{rccccc}
\hline
                 & $r_c$  & $\gamma^\dagger$   &$\gamma$  &  $z_s$    & $\delta$        \\
\hline
I-K              & 1.5    & 565                 &588$\pm3$ &  2.40     &    0.03$\pm0.05$ \\
I-K              & 2.0    & 577                 &611$\pm5$ &  2.38     &    0.02$\pm0.05$ \\
H                & 1.5    & 588                 &588$\pm3$ &  2.30     &    0.10$\pm0.05$ \\
Ref\cite{vega07} & 1.3    & -                   &602       &   -       &       -         \\
Ref\cite{chen07} & 0.98   & -                   &567       &   -       &       -         \\
\hline
\hline
\end{tabular}
\caption{\small Surface tension (bar nm) as derived from the integral of the surface tension profile ($\gamma^\dagger$) and from the global pressure tensor components ($\gamma$), position of the surface of tension $z_s$ (nm), and Tolman length $\delta$ (nm). The quantities have been calculated using the Irving-Kirkwood (I-K) and Harasima (H) paths, for different values of the cutoff radius $r_c$.}
\end{table}

In order to show the possibilities opened by using the Harasima
path and its feature of associating a pressure contribution to each
particle, we performed a decomposition of the lateral pressure
profile in the contributions coming from successive molecular layers,
starting from the surface one. To perform the layer-by-layer analysis,
we used the ITIM algorithm\cite{partay2008new,sega2013generalized},
extending its analysis capabilities to the extended GROMACS trajectory
format that includes virial contributions (this software is also
freely available, at \texttt{marcello-sega.github.io/gitim/}).
The interfacial analysis was performed on a molecular basis (i.e.~if
at least one atom in the molecule is at the interface, the complete
molecule is considered to be an interfacial one) using a probe
sphere radius of 0.2 nm. Vapour phase molecules were identified
using a neighbor list cluster-search algorithm based on a simple
oxygen-oxygen distance cutoff of 0.35 nm (corresponding to the first
minimum of the oxygen-oxygen radial distribution function of water).
A simulation snapshot, with molecules from different molecular
layers highlighted in different colors, as well as molecules in the
vapour phase as identified by the cluster-search, is reported in
Fig.\ref{snap}.

\begin{figure}[t]
\includegraphics[width=\columnwidth, trim=0 2 20 0,clip]{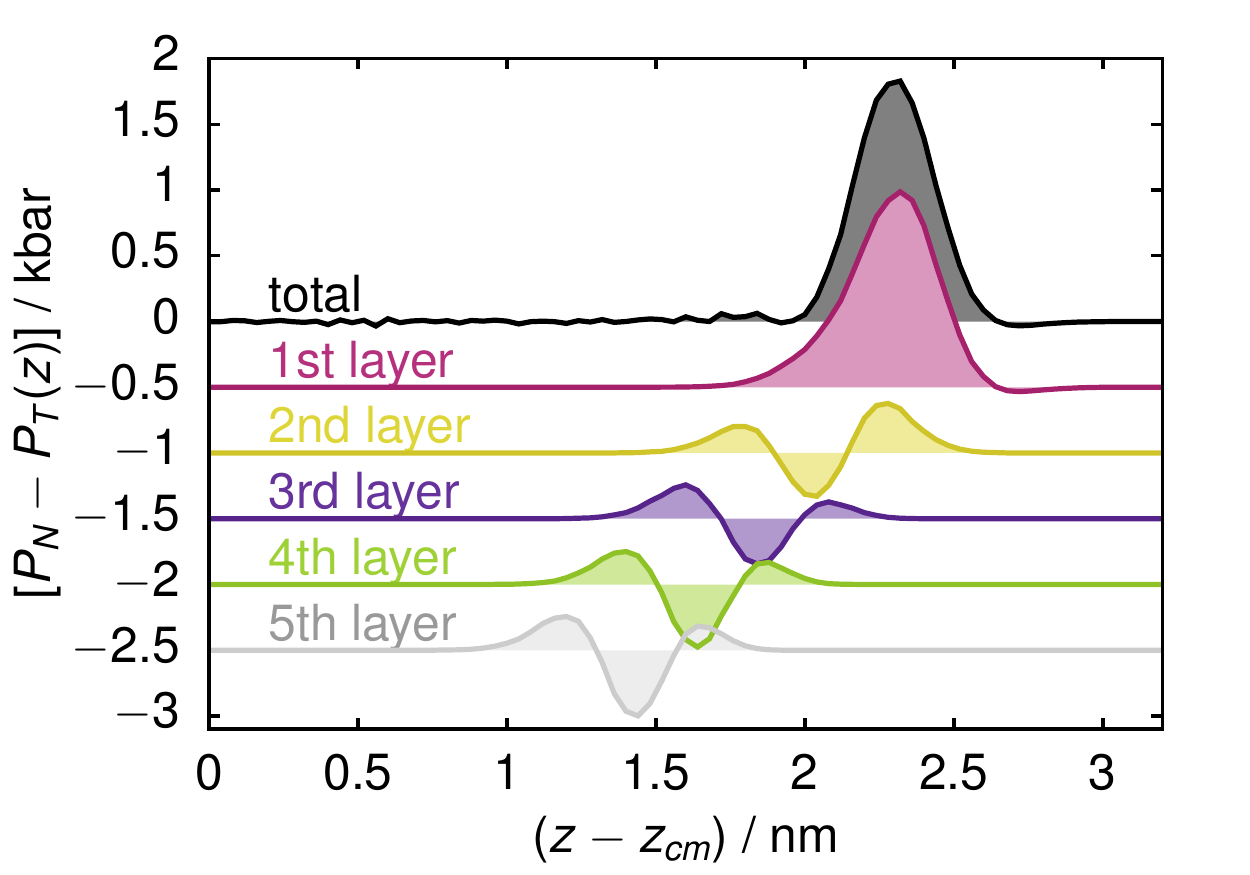}
\caption{Lateral pressure profile ($P_N-P_T(z)$) of the water/vapour interface. The contribution of each of the first 5 layers is reported (shifted along the vertical axis by 0.5 kbar each), using the same color scheme as for the simulation snapshot, Fig.\ref{snap}. The total profile is the curve at the top.\label{nonintr}}
\end{figure}
\begin{figure}
\includegraphics[width=\columnwidth,trim=0 0 20 0,clip]{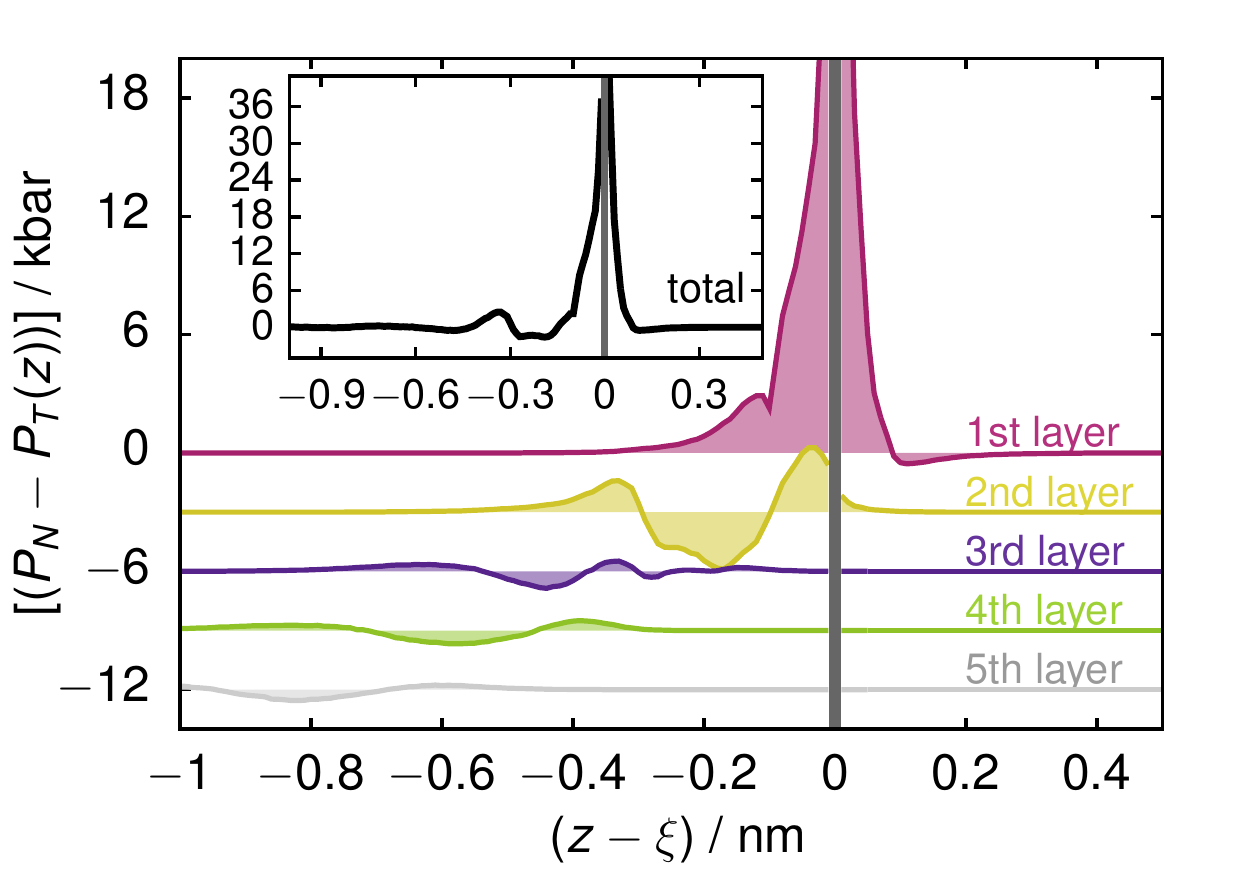}
\label{intrinsic}
\caption{Intrinsic lateral pressure profile ($P_N-P_T(z)$) of the water/vapour interface. The contribution of each of the first 5 layers is reported (shifted along the vertical axis by 3 kbar  each), using the same color scheme as for the simulation snapshot, Fig.\ref{snap}. The total profile is the curve in the inset. Delta-like contributions at the interface ($z=\xi$) are replaced by a gray band in the corresponding bin.\label{intr}}
\end{figure}

The presence of successive layers below the surface one is not
just a feature encountered in solids, but is also present in liquids,
and can be clearly seen by calculating the density profile in a local reference frame located at the interface, effectively removing the smearing introduced by the presence of capillary waves (which is more important the larger the allowed wavelengths in the simulation box are)\cite{rowlinsonwidom,chacon2003intrinsic}. If the interface position is  $z=\xi(x,y)$, the intrinsic density profile can then be defined in terms of the position $(x_i,y_i,z_i)$ of the $i$-th particle as
\begin{equation}
\rho_I(z)  = \frac{1}{A}\left\langle \sum_i \delta\left(z-z_i+\xi(x_i,y_i)\right)\right\rangle,
\end{equation}
where $A$ is the simulation box cross-sectional area along the macroscopic interface normal vector.
Similarly, we define the intrinsic pressure profile as

\begin{equation}
\gamma_I(z)= P_N - \frac{V}{A}\left\langle \sum_i P_T^i \delta\left(z-z_i+\xi(x_i,y_i)\right)\right\rangle.
\label{eq:intrinsic_gamma}
\end{equation}

Having access to the virial contribution of all particles, it is
straightforward to compute the pressure profile,
both with respect to the distance from the center of mass, (non-intrinsic
profile) and with respect to the local position of the interface
(intrinsic profile). In Fig.\ref{nonintr}, the total (top curve)
value of the profile of $\gamma(z)$ is reported, together with its
decomposition into the successive layers contributions, up to the
fifth layer.
It is important to notice that while the total normal pressure
profile must be a constant to ensure mechanical stability, this
is not true for the separate layer contributions. As we do not have direct
access to the normal component using the Harasima path, the layer
decomposition has to be understood, strictly speaking, as that of
the lateral pressure, which we offset by the value $P_N$ for
convenience. Nevertheless, this decomposition is still instructive,
as one can find that the first layer profile contributes the most
to the total profile, even though the latter is more narrowly
distributed than the former. The net contribution of each of the
layers far from the interface (i.e., the integral of the surface
tension distribution) is close to zero, consistently with the fact
that in this region the fluid is homogeneous and isotropic. The
distribution itself, however, is far from being identically zero,
and reflects the fact that in the oscillating layer, atoms are
undergoing different stresses depending on their distance from the
average position.

The width of the contribution
of the different layers is, on the other hand, quite constant and
rather close to the size of the first water coordination shell (of
radius 0.35 nm). Note that the distribution is normalized such that
its integral, from the center of the liquid slab ($z=0$) up to the box edge length, yields the surface tension. The intrinsic profile $\gamma_I(z)$, Fig.~\ref{intr}, gives
another point of view on the pressure profile distribution. The
intrinsic profile is calculated as a function of the distance from
the local surface position, $\xi(x,y)$, according to Eq.(\ref{eq:intrinsic_gamma}). In this way, the
effect of thermal capillary waves, which are corrugating the
surface, is removed. Note that the delta-like contribution at the
surface has been removed, and the corresponding bin blanked.  From
a comparison between the two plots, one can appreciate the considerable
narrowing of the pressure distribution, which decreases from 0.5
nm in the non-intrinsic case, to 0.3 nm in the intrinsic case,
clearly showing the smearing effect of capillary waves. A striking
difference between the pressure profile (both intrinsic and not)
of a molecular liquid such as water, and that of an atomic liquid like
argon\cite{sega2015layer}, is that the layers' contributions are
much less separated. Since the kissing condition between the probe
sphere and a surface atom takes into account the atomic radius,
some (point-like) hydrogen atoms in the second layer will be located
further than an oxygen atom of the first layer, contributing to the
enhanced layer interpenetration.  In addition, since we are considering molecular layers,
the contribution to the first layer comes also from atoms that are not right at the surface and, therefore, the first layer contribution is not just a delta function, as it is typical for simple
liquids\cite{chacon2003intrinsic}. The second, qualitative difference
with respect to simple liquids, is that the major part of the lateral
tension arises from atoms located in a narrow interval of 0.3 nm
around the interfacial centres, a range which is comparable with
one molecular diameter. The first molecular layer, in particular,
contributes to more than 90\% of the total surface tension. In simple
liquids, the distribution of the lateral pressure extends to almost
two atomic diameters, and the contribution of the first layer is
slightly less than 80\%.

\section{Computational Complexity\label{sec:scaling}}
\begin{figure}[t]
\includegraphics[width=\columnwidth,trim=0 0 20 0,clip]{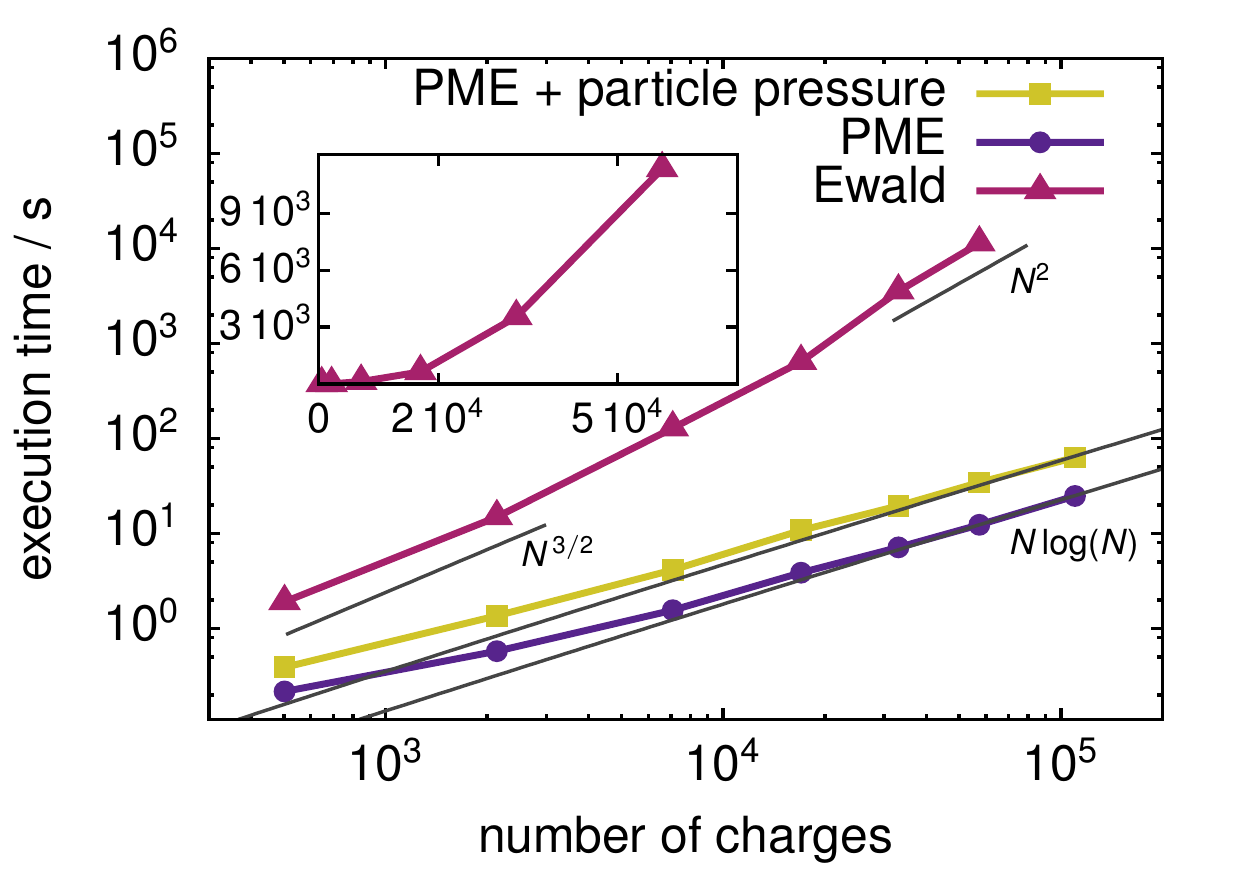}
\caption{Time needed to integrate 100 steps for systems of different sizes, from about 500 to more than $10^5$ water molecules. Circles and squares refer to the sPME execution time without and with local pressure calculation, respectively. Triangles represent the execution time when a plain Ewald method is used. The solid thin lines are the result of a fit to the function $N\log(N)$. The typical scaling behaviors for the plain Ewald calculation $N^{3/2}$ and  $N^2$ are also shown. Inset: the Ewald scaling in linear scale.\label{scaling}}
\end{figure}
The present algorithm for the calculation of the reciprocal space virial contributions
follows closely that of the calculation of forces in mesh-Ewald methods. Therefore, the scaling
is of the type $N\log(N)$. Regarding the prefactor, it is easy to make an estimate of the relative computational load with respect to the case where only forces are being computed. The functions $ f(\mathbf{m}^2)$ and $g_{\mu\nu}(\mathbf{m})$ are
usually being pre-calculated in order to provide energy and forces, and  this fact means that for the pressure profile computation, the additional calculations that need to be performed are one additional inverse FFT per tensor
element, and the corresponding back-interpolation. In terms of computational cost, the calculation of each of the tensor elements
has the same load as the force calculation in the sPME method (which requires only one FFT, at a difference with the PME method, which requires three).
Since we calculate only the diagonal elements of the virial
tensor, the theoretical prefactor is precisely 4 times larger than
for a standard sPME calculation. This is, however, the scaling one
would expect for the reciprocal part of the mesh-based algorithm
alone. As a matter of fact, van der Waals interactions, as well as
the short-range part of the electrostatic potential, are always
present in a typical simulation, with the result that in a real simulation setup,
this factor of four represents only an upper limit of the measured scaling.
To test the actual performance of the code, we ran a series of
simulations of bulk water at a density of about 1 g / cm$^3$ with
different box sizes, with a water content ranging from about 500
to more than $10^5$ water molecules, with the simulation protocol
and parameters already used for producing the pressure profiles, but a cut-off radius of 0.9 nm.
In Fig.\ref{scaling}, we report the time required to perform 10
integration timesteps, with and without calculating the local
pressure contribution. For comparison, the timings of runs using
the plain Ewald method are also reported. The result of a fit to a
$N\log(N)$ scaling gives, in the large $N$ limit, the prefactors
are $5.1$ and $1.9$ $\mu$s/timestep, with and without pressure
calculation, respectively. The simulations including the calculation
of the local pressure with sPME are in this case 2.7 times slower
than without including the pressure calculation.  It should be
noted, however, that in most practical cases it is not necessary
to compute the local pressure at every timestep and, therefore,
the method would introduce only a little overhead with respect to a
regular simulation.

\section{Conclusions\label{conclusions}}
We presented a method to calculate the long-range contribution of
the electrostatic interaction to the local pressure tensor, based
on mesh-based algorithms. We implemented our approach in the GROMACS
molecular simulation package and have tested it on a model of
water/vapour interface.  We have calculated the lateral pressure
profile with our approach, and compared it to the one computed using
the Irving-Kirkwood path, reporting also the location of the surface
of tension for the different models.  We have also shown how it is
possible to take advantage of the Harasima path decomposition to
calculate the layer-by-layer contribution to the intrinsic and
non-intrinsic lateral pressure profile.  The present method is
particularly efficient, as it displays the same scaling (in this
case, $N \log(N)$) of the algorithm used to compute long-range
correction to forces and energy.

This project is supported by the Hungarian OTKA Foundation under
project No. 109732, by the Action Austria Hungary Foundation under
project No. 93öu3104 and through ETN COLLDENSE, Grant H2020-MSCA-ITN-2014
No. 642774.  The computational results presented have been achieved
in using the Vienna Scientific Cluster (VSC). We thank M. McCaffrey for
proofreading the manuscript.

\providecommand{\latin}[1]{#1}
\providecommand*\mcitethebibliography{\thebibliography}
\csname @ifundefined\endcsname{endmcitethebibliography}
  {\let\endmcitethebibliography\endthebibliography}{}


\end{document}